# The omega-square hypothesis

## G. Molchan

*Institute of Earthquake Prediction Theory and Mathematical Geophysics,*

*84/32, Profsoyuznaya  Str., Moscow,117997, Russian Federation.*

*E-mail: molchan@mitp.ru*

**Summary**. The omega-square hypothesis assumes that the ground displacement $u(t)$ in the far-field zone decays as the inverse square of frequency in the range ~1-30 Hz.  This empirical fact remains theoretically unjustified.  Our analysis of the problem is based on an integral representation of $u(t)$ in terms of the source time function $f$  and on the spectrum analysis of local features in $f$ .  The goal is to select the local features that will enable one to generate the omega-square behavior of $u(t)$ on a large set of receivers. We found two appropriate fragments of $f$ :  first ,  $f$  exhibits a local inverse-square-root behavior near the rupture front where the frontal surface is piecewise smooth (but not  smooth   and not  rough  ); and second,  $f$  is bounded near the rupture front where the frontal surface has a slight roughness (the Hurst parameter near 1). These facts can be useful for understanding the  $\omega^{-2}$ spectral behavior of the kinematic source models.

## .1 Introduction.

It is commonly thought that the ground displacement in the far-field zone decays as  $\omega^{-2}$  in the range ~1-30Hz.  This behavior is known as the  $\omega^{-2}$  hypothesis and its justification is still a problem to be solved. The interest in a theoretical substantiation of this hypothesis peaked in the 1980s (Aki and Richards, 1980). However, the interest does not wane, because the problem is important for modeling strong motions for engineering purposes (Gusev 2013, 2014).

Our analysis of the $\omega^{-2}$- hypothesis is based on the following integral representation of the far-field displacements at a receiver site $G_{rec}$ (for simplicity, the signal is considered in the scalar form):

$$u(t) = A\int_{\Sigma} f(\mathbf{g}, t - dist(\mathbf{g}, G_{rec})/c)d\Sigma \approx A\int_{\Sigma} f(\mathbf{g}, t - t_0 + <\mathbf{g}, \boldsymbol{\gamma}_{\Sigma}>/c))d\Sigma. \qquad (1)$$

(see Aki and Richards, 1980). Here $\Sigma$ is a broken area of the planar fault with coordinates $(g_1, g_2) = \mathbf{g}$, $f$ is a source time function, i.e., the local slip velocity $\Delta\dot{u}(\mathbf{g}, t)$, $\boldsymbol{\gamma}_{\Sigma}$ is the orthogonal projection of the hypocenter-to-receiver direction $\boldsymbol{\gamma}$ onto the rupture plane, $<\cdot, \cdot>$ is the scalar product, $c$ is wave velocity, $t_0 c = dist(hypocenter, G_{rec})$ , $d\Sigma = dg_1 dg_2$ is an element of $\Sigma$ and the $A$ factor combines constant coefficients, geometric spreading, and the wave radiation pattern for unit force.

In the framework of fracture mechanics, solutions $u(t)$ or $f(\mathbf{g}, t)$ are known in exceptional cases and, as a rule, for 2D rupture problem (i.e., the 1-D source model)(Kostrov,1964; Nielsen and Madariaga, 2003). Three-dimensional rupture problem require numerical computations that impede HF analysis (Madariaga, 1976; Madariaga et al., 1998). To justify the $\omega^{-2}$ hypothesis we follow the line of argument which goes back to Madariaga (1977) and Achenbach and Harris (1978). These authors assume that the slip velocity has a number of universal 'topological features' which are responsible for the high frequency generation. Proceeding on these lines, we need to choose the physically reasonable features that systematically create the $\omega^{-2}$ behavior with a large set of receivers. The last requirement is essential when we deal with the 3D rupture problem. Below we focus on a piecewise-smooth and fractal peculiarities of $f$ . We find among them some acceptable solutions and show the difficulties in the way of justifying the $\omega^{-2}$ hypothesis.

## 2 Piecewise-smooth functions $f(\mathbf{g}, t)$.

The Fourier transform of (1) is

$$\hat{u}(\omega) = A\int d\Sigma \int e^{i\omega t} f(\mathbf{g}, t - t_0 + <\mathbf{g}, \boldsymbol{\gamma}_{\Sigma}>/c)dt$$

$$= Ae^{i\omega t_0}\int d\Sigma \int e^{i\omega(s - <\mathbf{g}, \boldsymbol{\gamma}_{\Sigma}/c>)} f(\mathbf{g}, s)ds = Ae^{i\omega t_0}\hat{f}(-\boldsymbol{\gamma}_{\Sigma} c^{-1}\omega, \omega) \qquad (2)$$

where $\hat{f}(\mathbf{k},\omega)$ is the Fourier transform of function $f(\mathbf{g},t)$ that is extended as zero beyond its support.

In the spectral terms our problem looks as follows: describe analytical, but physically reasonable, features of piecewise smooth functions $f(\mathbf{g},t)$ with a bounded support $\Omega_f$ such that $\hat{f}(\mathbf{p}\omega),\omega \gg 1$ has the omega-square asymptotics for a large set of vectors $\mathbf{p} = (-\gamma_\Sigma / c, 1)$. In applications, to simplify the inverse source problem or to generate the signal $u(t)$ with the appropriate properties, the models like the $k$-squared model of Herrero& Bernard (1994) are used. They are based on the postulate that $\hat{f}(\mathbf{k},\omega) \approx const|\mathbf{k}|^{-2}$ for large $|\mathbf{k}|$. In this case the omega-square hypothesis is valid automatically and can be adapted to the desirable frequency range ~1-10Hz. ( Hisada ,2001).

Because the Fourier transform is a linear operator, we can analyze the high frequency (HF) asymptotics using local fragments of $f(\mathbf{x})$, i.e., functions $f(\mathbf{x}|\mathbf{x}_{0i}) = f(\mathbf{x})\varphi(\mathbf{x}|\mathbf{x}_{0i})$ such that $f(\mathbf{x}) = \sum f(\mathbf{x})\varphi(\mathbf{x}|\mathbf{x}_{0i})$. Here, $\varphi(\mathbf{x}|\mathbf{x}_{0i})$ is an auxiliary smooth function that is equal to 0 outside a small vicinity of $\mathbf{x}_{0i}$, $O_{x_{0i}}$, and equals 1 near $\mathbf{x}_{0i}$. This function does not distort $f(\mathbf{g},t)$ near $\mathbf{x}_{0i}$ and makes it vanish smoothly outside of $O_{x_{0i}}$. In this way the use of $\varphi$ allows us to ignore the boundary values of $f(\mathbf{g},t)$ on $O_{\mathbf{x}_0}$. The HF asymptotics of the integral

$$\int e^{i\omega<\mathbf{p},\mathbf{x}>} f(\mathbf{x})\varphi(\mathbf{x}|\mathbf{x}_0)d\mathbf{x} := \hat{f}(\mathbf{p}\omega|\mathbf{x}_0) \qquad (3)$$

is treated as the contribution of $\mathbf{x}_0$ into the HF asymptotics of $\hat{u}(\omega)$.

In this section we shall assume that $f(\mathbf{g},t)$ is a continuous piecewise smooth function inside of its support, $\Omega_f$. If $f(\mathbf{g},t)$ is smooth near an inner point $\mathbf{x}_0 = (\mathbf{g}_0, t_0)$ of $\Omega_f$, then , as is well known, the contribution of $\mathbf{x}_0$ into the HF asymptotics of $\hat{u}(\omega)$ is negligibly small. Therefore, the high frequency behavior of $\hat{u}(\omega)$ is determined by the singularities of $f(\mathbf{g},t)$ inside of $\Omega_f$, by properties of $f(\mathbf{g},t)$ near the boundary of $\Omega_f$, and by properties of the boundary $\partial\Omega_f$ itself ($\partial\Omega_f$ will be termed the *frontal surface of the source*).

In what follows the support $\Omega_f$ and its boundary $\partial\Omega_f$ near a boundary point $\mathbf{x}_0 \in \partial\Omega_f$ will be specified by the following relations:

$$\Omega_f : \varepsilon t > \varepsilon t_r(\mathbf{g}|\mathbf{x_0}) \qquad\qquad \partial\Omega_f : t = t_r(\mathbf{g}|\mathbf{x_0}) \qquad\qquad (4)$$

where $\varepsilon = 1$ or $\varepsilon = -1$ .

The quantity $t_r(\mathbf{g}|\mathbf{x_0})$ can be treated as the rupture /healing time at g. In turn,

$$t_a(\mathbf{g}) = t_0 - <\mathbf{g}, \boldsymbol{\gamma}_\Sigma / c > + t_r(\mathbf{g}|\mathbf{x_0}) \qquad\qquad (5)$$

is the arrival time of the signal from the rupture front at $\mathbf{g}$ to the receiver point $G_{rec}$ .

Now we consider the contributions of some singularity points of $f$ .

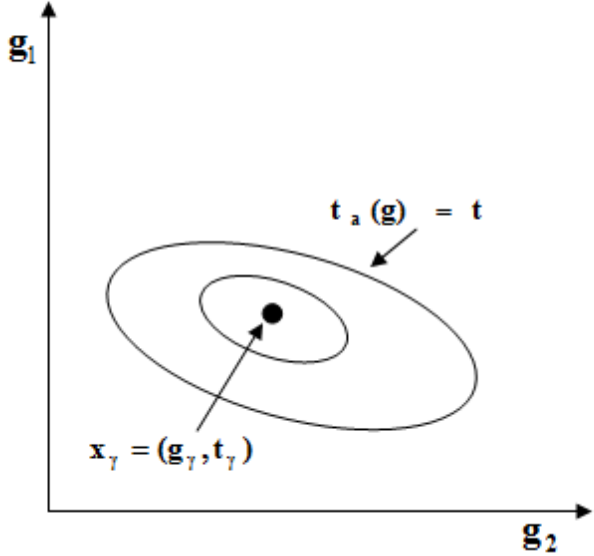

**Figure 1.** Isochrones of the time-arrival function $t_a(\mathbf{g})$ (the supersonic case (6)):

$\mathbf{x}_\gamma = (\mathbf{g}_\gamma, t_\gamma) \in \partial\Omega$ is the point where the frontal surface is smooth, the slip velocity is finite, and

$t_a = t_0 - <\mathbf{g}, \boldsymbol{\gamma}_\Sigma / c > + t$ has a local extreme on $\partial\Omega$ .

## 2.1 A smooth fragment of $f$ on $\partial\Omega_f$.

Let a surface $L$ be a smooth fragment of $\partial\Omega_f$ , $\mathbf{x}_0 \in L$ , $f$ is a smooth function near $\mathbf{x}_0$ up to the boundary $\partial\Omega_f$ where $f(\mathbf{x}) \neq 0$ . This type of $f$ -behavior is typical for the healing phase of rupture dynamics (an example see in Madariaga et al., 1998). The condition $f(\mathbf{x}_0) \neq 0$ implies the discontinuity $f$ near $\mathbf{x}_0$ as a function in the whole space.

To find the HF contribution of $\mathbf{x}_0$, we can use a general fact related to the HF asymptotics of smooth functions in a domain with a smooth boundary (see (4.21) in Fedoryuk, 1987).

Assume that $L$ and a plane $t - <\mathbf{g}, \boldsymbol{\gamma}_\Sigma> / c >= const$ have an isolated tangent point $\mathbf{x}_\gamma = (\mathbf{g}_\gamma, t_\gamma)$ such that the total (Gaussian) curvature of $t_r(\mathbf{g}|\mathbf{x_0})$ at $\mathbf{x}_\gamma$ is non-zero, $K_\gamma \neq 0$. In other words, $\mathbf{x}_\gamma$ is a *regular stationary point* of the time arrival function $t_a(\mathbf{g})$, (see Fig1).

Then the contribution of $\mathbf{x}_0$ into the HF asymptotics of $\hat{u}(\omega)$ is

$$\hat{u}(\omega \mid \mathbf{x}_0) \approx \omega^{-2} \cdot 2\pi |A| \left\| K_\gamma \right\|^{-1/2} (1 + \left| \boldsymbol{\gamma}_\Sigma / c \right|^2)^{-1} f(\mathbf{x}_\gamma) \varepsilon_\gamma \delta_{\mathbf{n}_\gamma}, \quad \left| \varepsilon_\gamma \right| = 1, \quad \left| \delta_{\mathbf{n}_\gamma} \right| = 1, \quad \omega \to \infty, \qquad (6)$$

where $\varepsilon_\gamma$ depends on $\mathbf{g}_\gamma$; $\delta_{\mathbf{n}_\gamma}$ depends on the outer normal $\mathbf{n}_\gamma$ to $L$ at $\mathbf{x}_\gamma$. In addition

$$\delta_{-\mathbf{n}_\gamma} = \overline{\delta}_{\mathbf{n}_\gamma}, \text{ and } \text{Re } \delta_{\mathbf{n}_\gamma} = 0 \text{ for } K_\gamma < 0. \qquad (7)$$

If $\mathbf{x}_\gamma$ does not exist, the HF contribution of $\mathbf{x}_0$ is negligibly small.

## 2.2 A piecewise smooth fragment of $f$ within $\Omega_f$.

The last statement can be applied to the case where $f$ looses its smoothness on a smooth 2-D surface $L$ within $\Omega_f$

Suppose that $\mathbf{x}_0 \in L \subset \Omega_f$, $f$ is continuous in a vicinity $O_{x_0}$ of $\mathbf{x}_0$, where $f(\mathbf{x}) \neq 0$; $f$ is smooth in $O_{x_0} \setminus L$ and in $L$. The surface $L$ divides $O_{x_0}$ into two parts $O_{x_0}^\pm$. Interpreting two sides of $L$ as fragments of the boundary $\partial \Omega_f$, we can apply statement (5) to the domains $O_{x_0}^\pm$ and sum the results. Using (7), we will get the contribution of $\mathbf{x}_0$ into the HF asymptotics of $\hat{u}(\omega)$. It is given by the same formula (6) with $\delta_{\mathbf{n}_\gamma}$ replaced with $2\text{Re }\delta_{\mathbf{n}_\gamma}$. The last term is zero for $K_\gamma < 0$.

Thus, under some additional conditions, a 2-D singularities of $f$ within $\Omega_f$ can generate the $\omega^{-2}$ behavior of $\hat{u}(\omega)$ as well.

## 2.3 Existence of the stationary points $\mathbf{x}_\gamma$.

At the tangent point $\mathbf{x}_\gamma = (\mathbf{g}_\gamma, t_\gamma)$ of $L = \partial\Omega_f$ and the plane $t - <\mathbf{g}, \boldsymbol{\gamma}_\Sigma / c >= const$ the gradient $\nabla L$ is proportional to the vector $(1, -\boldsymbol{\gamma}_\Sigma / c)$. Therefore, if $\partial\Omega_f$ is given by equation (4):

$t = t_r(\mathbf{g} | \mathbf{x}_0)$ , we must have

$$\nabla t_r(\mathbf{g}_\gamma | \mathbf{x}_0) = \boldsymbol{\gamma}_\Sigma / c \ . \tag{8}$$

Since $\left|\nabla t_r(\mathbf{g}_\gamma)\right|$ is the local slowness, $1/v_r(\mathbf{g}_\gamma)$, of the frontal surface at $\mathbf{g}_\gamma$, and $\left|\boldsymbol{\gamma}_\Sigma\right| \leq 1$, one has $v_r(\mathbf{g}_\gamma) \geq c$ . In other words, the front velocity at $\mathbf{x}_\gamma = (\mathbf{g}_\gamma, t_\gamma)$ is supersonic, that is rarely observed in earthquakes (Rosakis et al., 1999; Madariaga et al, 2000).

Thus, a smooth 2D singularity (including $\partial\Omega_f$), where $f$ or $\nabla f$ are discontinuous, cannot be a regular source of the $\omega^{-2}$- behavior of the displacement spectra.

## 2.4 Unbounded $f$ near $\partial\Omega_f$.

The slip source velocity function is unbounded near the crack edge. The classical self-similar circular shear crack model by Kostrov (1964) specifies this as follows:

$$f(\mathbf{g}, t) = A \cdot d / dt(v_r^2 t^2 - |\mathbf{g}|^2)_+^{1/2}, \qquad v_r < c \bullet \tag{9}$$

Below, the notation $(a)_+$ means $a$ for positive $a$ and 0 otherwise. Therefore $\partial\Omega_f$ in (9) is a cone, $\{t = |\mathbf{g}| / v_r\}$.

A more general local structure of $f$ near $\partial\Omega_f$ is known from the 2D models of an arbitrarily moving crack. According to the models of Madariaga (1977, 1983), $f$ near the crack tip can be represented as follows:

$$f(g, t) = l'(t)(l(t) - g)_+^{-1/2} V(g, t) \ , \tag{10}$$

where $l(t)$ is the current position of the crack tip. According to Madariaga (1977 , 1983), if the rupture velocity changes abruptly, the singularity (10) generates the $\omega^{-2}$ behavior in the near field zone. Under the same condition, the far field displacement spectra will have a non-standard order, namely $O(\omega^{-3/2})$.

One more important example is considered by Madariaga et al. (2006) where the fault contains a single localized kink. As a result, the boundary $\partial\Omega_f$ in such a model is locally piecewise smooth.

These examples can be generalized as follows:

$$f(\mathbf{g},t) = A \cdot (t - t_r(\mathbf{g}))_+^{-1/2} V(\mathbf{g},t), \tag{11}$$

where $t_r(\mathbf{g})$ is the local representation of the piecewise frontal surface.

The following holds for the case of a smooth function $t_r(\mathbf{g})$ (see Appendix A2):

Suppose (11) holds in some vicinity of $\mathbf{x}_0 \in \partial \Omega_f$ with smooth components $t_r(\mathbf{g})$ and $V(\mathbf{g},t)$. If $t_a(\mathbf{g})$ has the regular stationary point $\mathbf{x}_\gamma$ (see Section **2.1**), then the contribution of $\mathbf{x}_0$ into the HF asymptotics of $\hat{u}(\omega)$ has the order $O(\omega^{-3/2})$. Otherwise ($\mathbf{x}_\gamma$ does not exist), the contribution is negligibly small.

In the virtue of Section **2.3**, the singularity (11) with smooth components $t_r(\mathbf{g})$ and $V(\mathbf{g},t)$ cannot be a regular source of the power-type HF behavior of $\hat{u}(\omega)$.

The Kostrov model (9) is an example of (11) with smooth components. In this model the HF asymptotics of $\hat{u}(\omega|\mathbf{x}_0)$, $\mathbf{x}_0 \neq 0$ is negligibly small because the rupture velocity is subsonic. Note, in addition, that the curvature of $\partial \Omega_f$ is zero in this example.

## 2.5 A piecewise smooth fragment of $\partial \Omega_f$ in (11).

Consider the local model (11) with a smooth $V(\mathbf{g},t)$ and a continuous, piecewise smooth surface $t_r(\mathbf{g})$ near $\mathbf{x}_0 = (\mathbf{g}_0, t_0) \in \partial \Omega_f$. More precisely, $t_r(\mathbf{g})$ is represented in a vicinity of $\mathbf{x}_0 \in \partial \Omega_f$ by two smooth surfaces $t_r^\pm(\mathbf{g})$, $g \in O_{\mathbf{g}_0}^\pm$ where the domains $O_{\mathbf{g}_0}^\pm$ are not necessarily different. These surfaces intersect along a smooth 1-D curve $L$, (see Fig.2). The relation

$$\hat{L}: \ t_r^+(\mathbf{g}) - t_r^-(\mathbf{g}) = 0 \tag{12}$$

gives the orthogonal projection of $L$ onto the fault plane $\{(g_1, g_2)\}$. For simplicity we suppose that $\hat{L}$ is given by the relation $g_2 = \psi(g_1)$ with a smooth function $\psi(\cdot)$.

In the previous sections the power-law HF asymptotics of $\hat{u}(\omega)$ was defined by the isolated stationary regular points $\mathbf{x}_\gamma = (\mathbf{g}_\gamma, t_\gamma)$ of the time-arrival function $t_a(\mathbf{g})$. The subsonic restriction, $v_r(\mathbf{g}_\gamma) < c$, excludes the existence of the critical point $\mathbf{x}_\gamma = (\mathbf{g}_\gamma, t_\gamma)$.

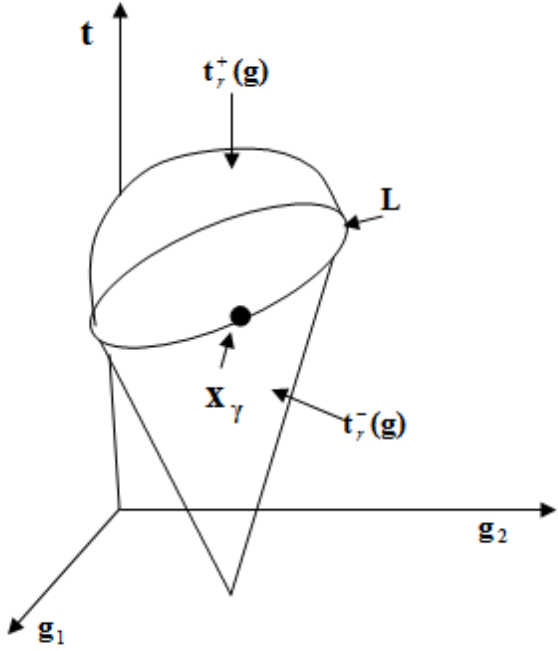

**Figure 2.** An illustration of statement (15): $L$ is a 1-D singularity of the frontal surface $\partial \Omega_f = t_r^+ \cup t_r^-$; $\mathbf{x}_\gamma = (\mathbf{g}_\gamma, t_\gamma) \in L$ is the point where $t_a = t_0 - <\mathbf{g}, \boldsymbol{\gamma}_\Sigma / c > + t$ has a local extreme on $L$, and the slip velocity has the inverse square root singularity on $\Omega_f$ (see (11)).

In the situation considered here, the analogue of $\mathbf{g}_\gamma$ is the *conditional stationary point* $\widetilde{\mathbf{g}}_\gamma$ of the function $t_a(\mathbf{g})$ contracted on $\hat{L}$, i.e., $\widetilde{\mathbf{g}}_\gamma = (\widetilde{g}_{1\gamma}, \psi(\widetilde{g}_{1\gamma}))$ and $\widetilde{g}_{1\gamma}$ is the stationary point of $t_a(g_1, \psi(g_1)) := S(g_1)$. By definition, the point $\widetilde{\mathbf{g}}_\gamma$ is *regular* if the following relations hold:

$$S'(\widetilde{g}_{1\gamma}) = 0, \qquad \widetilde{K}_\gamma := S''(\widetilde{g}_{1\gamma}) \neq 0. \tag{13}$$

The critical point $\widetilde{\mathbf{g}}_\gamma$ can also be found as a solution to any of the following two equations:

$$\nabla t_r^\pm(\mathbf{g}|\mathbf{x}_0) - \hat{\boldsymbol{\gamma}}_\Sigma / c = C_\pm \nabla \hat{L}(\mathbf{g}), \quad \mathbf{g} \in L, \tag{14}$$

where $C_\pm$ are unknown constants, $\hat{L}(\mathbf{g})$ is a smooth function, and $\hat{L}(\mathbf{g}) = 0$ is the equation for $\hat{L}$. By (12), the solutions of (14) are identical for the different symbols (+,-).

Now we are ready to formulate a statement that is proved in Appendix A3.

Suppose that (11) holds in some vicinity of $\mathbf{x}_0 \in \partial\Omega_f$ with the components $t_r(\mathbf{g})$ и

$V(\mathbf{g},t)$ described above. If the conditional stationary point $\widetilde{\mathbf{g}}_\gamma$ exists and is regular, then the

contribution of $\mathbf{x}_0$ into the HF asymptotics of $\hat{u}(\omega)$ is

$$\hat{u}(\omega \mid \mathbf{x}_0) \approx \omega^{-2} \cdot |A| \varepsilon \sqrt{2\pi} \left| \widetilde{K}_\gamma \right|^{-1/2} V(\widetilde{\mathbf{g}}_r, t_r(\widetilde{\mathbf{g}}_\gamma)) \Delta_\gamma, \quad |\varepsilon| = 1, \tag{15}$$

where

$$\Delta_\gamma = [\partial/\partial g_2 t_r^+(\widetilde{\mathbf{g}}_\gamma) - \hat{\gamma}_\Sigma^{(2)}/c]^{-1} - [\partial/\partial g_2 t_r^-(\widetilde{\mathbf{g}}_\gamma) - \hat{\gamma}_\Sigma^{(2)}/c]^{-1}. \tag{16}$$

The result (15) is correct if both terms in (16) are finite. Since $\nabla t_r^\pm(\mathbf{g}|\mathbf{x}_0) \neq \boldsymbol{\gamma}_\Sigma/c$ (see Section

**2.3**), this restriction is easily overcame. Indeed, if $\Delta_\gamma = \infty$, we have to use a specific representation of $\hat{L}$ for each of the domains $O_{\mathbf{g}_0}^\pm$ in the form $g_2 = \psi(g_1)$ or $g_1 = \widetilde{\psi}(g_2)$. As a result, (15,16) will contain two terms with different $\widetilde{K}_\gamma := S''(\widetilde{g}_{1\gamma})$ and the corresponding finite component of $\Delta_\gamma$.

  If $\widetilde{\mathbf{g}}_\gamma$ does not exist, then the HF contribution due to $\mathbf{x}_0$ is negligibly small.

**Example 1.** In addition to Fig.2, we illustrate the previous statement on the model by Sato and

Hirasawa (see Aki and Richards, 1980). In fact this model is an abruptly stopped Kostrov crack

model (9) with the slip function

$$\Delta u(\mathbf{g},t) = A[[\min(vt, r_0)]^2 - |\mathbf{g}|^2]_+^{1/2}. \tag{17}$$

In this case $\Omega_f$ is a cone $\{t > |\mathbf{g}|/v\}$ with a planar base $B = \{|\mathbf{g}| \leq r_0, t = r_0/v\}$. The boundary $\partial\Omega_f$

is given by two smooth functions

$$t_r^+(\mathbf{g}) = r_0/v, \mathbf{g} \in O^+ = \{|\mathbf{g}| \leq r_0\} \text{ and } t_r^-(\mathbf{g}) = |\mathbf{g}|/v, \mathbf{g} \in O^+ \setminus \{0\}. \tag{18}$$

The line $\hat{L}$: $t_r^+(\mathbf{g}) - t_r^-(\mathbf{g}) = 0$ is a circle $\{|\mathbf{g}| = r_0\}$; the contraction of $t_a(\mathbf{g})$ on $L$ is a linear function: $S(\mathbf{g}) = const - <\mathbf{g}, \hat{\boldsymbol{\gamma}}_\Sigma/c>, \mathbf{g} \in \hat{L}$. Therefore $S(\mathbf{g})$ has two extremes at points

$\widetilde{\mathbf{g}}_\gamma = \pm\hat{\boldsymbol{\gamma}}_\Sigma/|\hat{\boldsymbol{\gamma}}_\Sigma| \cdot r_0$ (conditional stationary points in our terminology). Therefore for any direction

$\hat{\boldsymbol{\gamma}}_\Sigma$ there exists $\mathbf{x}_0 = (\mathbf{g}_0 = \pm\hat{\boldsymbol{\gamma}}_\Sigma/|\hat{\boldsymbol{\gamma}}_\Sigma| \cdot r_0, t_0 = r_0/v)$ such that $\hat{u}(\omega \mid \mathbf{x}_0) = O(\omega^{-2})$.

**2.6 Conditional stationary points of $t_a(\mathbf{g})$.**

The value of the $\omega^{-2}$-asymptotics (15) depends on the solubility of (14) for a large set of vectors $\hat{\boldsymbol{\gamma}}_\Sigma$. Let us consider this question.

Madariaga (1977) gave some arguments in favor of the power law asymptotics for the 3D problem when the rupture velocity field jumps simultaneously on a segment of the rupture front. In this special case the frontal surface looses its smoothness on the line $\hat{L}$: $t_r(\mathbf{g}) = t_0$ and therefore (14) looks as follows:

$$\nabla t_r^\pm(\mathbf{g}|\mathbf{x}_0) = C^\pm \hat{\boldsymbol{\gamma}}_\Sigma \; , \quad \mathbf{g} \in \hat{L} \; . \tag{19}$$

All vectors $\hat{\boldsymbol{\gamma}}_\Sigma$ form the unit disk $B$. Consider the largest angle $\alpha$ between two vectors $\nabla t_r^+(\mathbf{g}|\mathbf{x}_0)$, $\nabla t_r^+(\mathbf{g}'|\mathbf{x}_0)$ where $\mathbf{g}$, $\mathbf{g}'$ are arbitrary points on $\hat{L}$ ($\alpha = \pi$ in the model from Example 1). Obviously, the set of $\hat{\boldsymbol{\gamma}}_\Sigma$ for which (19) is solvable forms the two-sided circular sector of angle $\alpha$ in $B$. The greater the curvature of $\hat{L}$ near $\mathbf{x}_0 \in \partial\Omega_f$ the larger is the set of permissible $\hat{\boldsymbol{\gamma}}_\Sigma$.

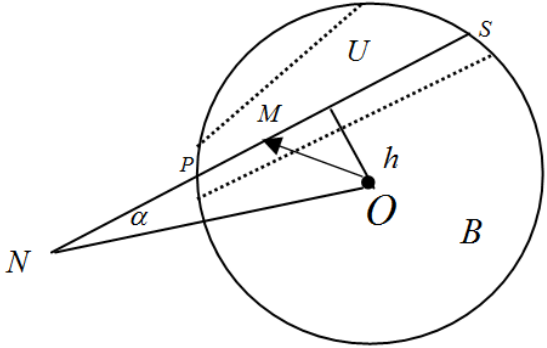

**Figure 3**. Equation (14): $[c\nabla t_r^+(\mathbf{g}|\mathbf{x}_0) := \mathbf{ON}] = [\hat{\boldsymbol{\gamma}}_\Sigma := \mathbf{OM}] + [C^+ c\nabla\hat{L}(\mathbf{g}) := \mathbf{MN}]$ ,

*Notation*: $B$ is the unite disk of all possible vectors $\hat{\boldsymbol{\gamma}}_\Sigma$; $\mathbf{OM}$ is one of $\hat{\boldsymbol{\gamma}}_\Sigma$ for which equation (14) is solvable ; the segment $PS$ is the set of $\hat{\boldsymbol{\gamma}}_\Sigma$ for which solution of (14) is the same as for $\mathbf{OM}$ ; the area between the dotted lines, $U$ , is a possible union of the $PS$-sets depending on the local variability of $\mathbf{MN}$; $h = |\mathbf{ON}|\sin\alpha = dist(O, PS)$ .

In the general case (see Fig3) the solution $\widetilde{\mathbf{g}}_\gamma$ of (14), if any, is common for all $\hat{\boldsymbol{\gamma}}_\Sigma$ belonging to a chord of $B$ that is at a distance $h = cv_r^{-1}(\widetilde{\mathbf{g}}_\gamma)\left|\sin(\alpha)\right|$ from the center of $B$. Here $\alpha$ is the angle between vectors $\nabla t_r^+(\widetilde{\mathbf{g}}_\gamma)$ and $\nabla\hat{L}(\widetilde{\mathbf{g}}_\gamma)$. Since $h \le \left|\hat{\boldsymbol{\gamma}}_\Sigma\right|$, one has

$$cv_r^{-1}(\widetilde{\mathbf{g}}_\gamma)\left|\sin(\alpha)\right| \le \left|\hat{\boldsymbol{\gamma}}_\Sigma\right| \quad . \tag{20}$$

We get an analogue of the supersonic condition that we found in section 2.3 for the unconditional stationary point of $t_a(\mathbf{g})$.

By (14), if the vector field $\nabla t_r(\mathbf{g})$ loses continuity locally, and weakly varies before the jump, then one comes again to the previous conclusion: the greater the curvature of $\hat{L}$ the larger is the set of $\hat{\boldsymbol{\gamma}}_\Sigma$ having the $\omega^{-2}$ behavior of the displacement spectra. This scenario with a single and small singularity of $\hat{L}$ cannot be stable to generate the $\omega^{-2}$ behavior on a large set of receivers. The other possibility is to admit the multiplicity and chaotic orientation of the $\hat{L}$-defects.

# 3. A rough frontal surface.

There is another, formally identical to (1), integral representation of the far-field displacement, where the source function $f$ is interpreted as a local stress drop and $\Sigma$ as an area occupied by the initial asperities (Kostrov and Das, 1988; Boatwright, 1988; Gusev, 1989). In this case the physical restrictions on $f$ are vague. Mikumo and Miyatake (1979) studied numerically the problem of random variation of stress drop over the fault. The fracture process was found to be quite chaotic, with no clearly distinguishable fracture front. This circumstance stimulated a stochastic or a fractal approach to kinematic rupture models.

Gusev (2014) proposed and analyzed numerically the following model

$$f = V(\mathbf{g})\phi(t - t_r(\mathbf{g})) , \tag{21}$$

where $\phi(t) \ge 0$ is a smooth function on the semi-axis $t \ge 0, \phi(0) \ne 0$ and $\phi(t) = 0$ outside of $[0, t_0]$; $V(\mathbf{g})$ is the local stress drop at a point $\mathbf{g}$ which is independent of time; $t_r(\mathbf{g})$ is the rupture time at point $\mathbf{g}$; $V(\mathbf{g})$ and $t_r(\mathbf{g})$ are continuous. The rupture is mostly unilateral with a nearly straight rupture front:

$$t_r(\mathbf{g}) = <\mathbf{g}, \boldsymbol{\gamma}_r> / v + \delta \tilde{t}_r(\mathbf{g}), \tag{22}$$

where $\boldsymbol{\gamma}_r$, $|\boldsymbol{\gamma_r}| = 1$, is the dominant direction of rupture front and $\delta \tilde{t}_r(\mathbf{g})$ is a small perturbation. In this model the support of $f$ is given by the following relation

$$\Omega_f : t_r(\mathbf{g}) < t < t_r(\mathbf{g}) + t_0 \tag{23}$$

Function $f$ is smooth in $t$ for inner points of $\Omega_f$. Using this fact, it is easy to show that the main contribution into the asymptotics of $\hat{u}(\omega)$ comes from the points of the frontal surface $\partial \Omega_f$. Moreover, it comes from the first boundary: $t = t_r(\mathbf{g})$, because $f$ as a function in whole space is discontinuous on the surface $t = t_r(\mathbf{g})$ only.

The main feature of the Gusev model is the fractality of $t_r(\mathbf{g})$ and $V(\mathbf{g})$, that is, their degree of smoothness is below 1. Fractal models of $f$ are simpler to study using the language of random processes. We shall say that a random function $\xi(\mathbf{g})$ has a smoothness of order $0 < H_\xi < 1$, if

$$E\left|\xi(\mathbf{g}) - \xi(\tilde{\mathbf{g}})\right|^2 \sim C(\mathbf{g})\left|\mathbf{g} - \tilde{\mathbf{g}}\right|^{2H_\xi}, \qquad \mathbf{g} - \tilde{\mathbf{g}} \to 0 , \tag{24}$$

where $E$ is the symbol of mathematical expectation, $C(\mathbf{g}) \neq 0$ is a smooth function; $H_\xi$ is known also as the Hurst parameter. The fractal condition (24) is local. It is more flexible compared with the power spectrum asymptotics that is commonly accepted for the slip distribution (Herrero&Bernard,1994).

The Gusev model treats $t_r(\mathbf{g})$ and $V(\mathbf{g})$ as random functions with the Hurst parameters $H_r$ and $H_V$, respectively. Note that for any smooth function that satisfies (24), we have $H = 1$. Considering a random signal, we shall speak of the root-mean-square (r.m.s.) spectra instead of individual spectra, $\hat{u}(\omega)$:

$$r.m.s.\hat{u}(\omega) = (E\left|\hat{u}(\omega)\right|^2)^{1/2} . \tag{25}$$

It is convenient to extend model (21) as follows:

$$f = V(\mathbf{g})\phi_\beta(t - t_r(\mathbf{g})) , \qquad \phi_\beta(t) = t_+^{\beta-1}\phi(t), \qquad \beta > 0 , \tag{26}$$

where $\phi(t)$ a smooth function, $\phi(0) \neq 0$, and $\phi(t) = 0$ for $t \geq t_0$. The Gusev model corresponds to $\beta = 1$. For (26),

$$\hat{f}(\widetilde{\mathbf{p}}\omega,\omega)=\hat{\phi}_{\beta}(\omega)\int e^{i(t_{r}(\mathbf{g})+<\widetilde{\mathbf{p}},\mathbf{g}>)\omega}V(\mathbf{g})d\Sigma \quad , \quad \widetilde{\mathbf{p}}=-\gamma_{\Sigma}/c \quad . \tag{27}$$

By the Erdelyi lemma (A1) , $\hat{\Phi}(\omega)=O(\omega^{-\beta})$ , $\omega>>1$ . The HF asymptotics of the integral term in (27) was studied by Molchan (2015) for different stochastic functions $t_{r}(\mathbf{g})$ and $V(\mathbf{g})$ . Combining these facts, one can immediately derive the results for the fractal source models.

### 3.1 Models of $\delta t_{r}(\mathbf{g})$ and $V(\mathbf{g})$ .

We consider $\delta t_{r}(\mathbf{g})=\xi^{2}(\mathbf{g})$ , $\mathbf{g}\in\Sigma$ , where $\xi(\mathbf{g})$ is a Gaussian random function such that $E\xi(\mathbf{g})=0$ and $E\xi(\mathbf{g})\xi(\widetilde{\mathbf{g}}):=B(\mathbf{g},\widetilde{\mathbf{g}})$ is a smooth function both outside of the diagonal D: $\mathbf{g}=\widetilde{\mathbf{g}}$ and on D. The smoothness of $\xi(\mathbf{g})$ is controlled by the Hurst parameter $0<H_{\xi}<1$ . The relation

$$\delta t_{r}(\mathbf{g})-\delta t_{r}(\widetilde{\mathbf{g}})=\xi^{2}(\mathbf{g})-\xi^{2}(\widetilde{\mathbf{g}})=(\xi(\mathbf{g})-\xi(\widetilde{\mathbf{g}}))(\xi(\mathbf{g})+\xi(\widetilde{\mathbf{g}})) \tag{28}$$

explains why we are led to think that the Hurst parameters for $\delta t_{r}(\mathbf{g})$ and $\xi(\mathbf{g})$ are identical. The additional restrictions on $\xi(\mathbf{g})$ are technical in character. They are specified by the requirements C1-C4 (Section 3.1) in (Molchan, 2015).

The random function $V(\mathbf{g})$ has smooth mean and variance; the covariance $EV(\mathbf{g})V(\widetilde{\mathbf{g}})$ is a smooth function outside of the diagonal D: $\mathbf{g}=\widetilde{\mathbf{g}}$ ; the smoothness of $V(\mathbf{g})$ is controlled by the Hurst parameter $0<H_{V}<1$ .

### 3.2 Fractal fragment of $f$ near $\partial\Omega_{f}$ .

We assume that (26) holds in some vicinity $O_{\mathbf{x}_{0}}$ of $\mathbf{x}_{0}=(\mathbf{g}_{0},t_{0})\in\partial\Omega_{f}$ ; the components of $f(\mathbf{g},t)$ , $V(\mathbf{g})$ and $t_{r}(\mathbf{g})$ , are independent random functions with the properties prescribed above (see Section **3.1** and relation (22)) .

Then the contribution of the vicinity of $\mathbf{x}_{0}$ into the asymptotics of $\hat{u}(\omega)$ is

$$r.m.s.\hat{u}(\omega\,|\,\mathbf{x}_{0})\approx\omega^{-(1+\beta+(1-H_{\xi})/2)}\cdot\left|A\right|\Gamma(\beta)C_{V,\xi}\left|\gamma_{\Sigma}/c-\gamma_{r}/v\right|^{-1+H_{\xi}/2}, \qquad \omega\to\infty . \tag{29}$$

The amplitude factor $C_{V,\xi}=C_{V,\xi}(O_{\mathbf{x}_{0}})$ depends on the behavior of the covariance functions both of $V(\mathbf{g})$ and $\xi(\mathbf{g})$ near the diagonal D: $\mathbf{g}=\widetilde{\mathbf{g}}$ only, i.e., $C_{V,\xi}$ is formed by the incoherent radiation of the front points and therefore $C_{V\xi}^{2}(O_{\mathbf{x}_{0}})$ is an additive function of the set $O_{\mathbf{x}_{0}}$ (see Molchan (2015), formula (27)).

Note that the limiting exponent $H_\xi = 1$ cannot be used in (29) because of the amplitude factor. This is clear a priori because the power law behavior of displacement spectra is unstable in the smooth case (see Section 2), namely, with the Hurst parameter equal to 1.

**Comments**

a) The asymptotic results for the fractal and the smooth cases are substantially different: in the fractal case the contribution of a boundary point $\mathbf{x}_0$ in the HF radiation depends on the entire vicinity $O_{x_0}$ and is nonzero. In the smooth case, the power law HF contribution of $\mathbf{x}_0$ depends on an isolated point and some additional analytical conditions. Therefore this contribution is unstable under changes in receiver position.

b) The power law exponent in (29), $\theta = 1 + \beta + \delta(H_r)$, as a function of the Hurst parameter of $\partial\Omega_f$, is not universal. However, the following property: $\delta(H_r) \to 0$ as $H_r \to 1$, is generic (Molchan, 2015). Therefore, to exclude any speculations relative to the power law index $\theta$, it is reasonable to discuss the quasi-regular case of $t_r(\mathbf{g})$, i.e., $H_r \sim 1$. In this case $\theta \approx 1 + \beta$, where $\beta = 1$ and $\beta = 1/2$ can be interesting.

If $\beta = 1$, the fragment of the source function is bounded near the frontal surface. In this case a slight roughness of $\partial\Omega_f$ implies the stable index $\theta \approx 2$. On the contrary, the smoothness of $\partial\Omega_f$ yields the same value, $\theta = 2$, but practically in one unrealistic situation only: the local front velocity is supersonic (see Section **2.1-2.3**).

The value $\beta = 1/2$ in (26) is reasonable for the crack propagation phase. In this case the piece-wise smooth fragment of $\partial\Omega_f$ can generate the $\omega^{-2}$ behavior of the far-field spectra at some observation points (see Section **2.5**). On the contrary, the slight roughness of $\partial\Omega_f$ leads to a stable, but unrealistic, index: $\theta \approx 3/2$. This means that the square-root behavior of the slip near the rupture front ( $\beta = 1/2$ ) and the slight roughness of $\partial\Omega_f$ are physically incompatible. It would be interesting to find some physical arguments for this. We recall that the same result, $\theta = 3/2$, we can have also with $\beta = 1/2$ and a smooth $\partial\Omega_f$. To reject this case we used the subsonic restriction.

## 4. Conclusions and Discussion.

The adequacy of any kinematic model of seismic radiation can be tested using dynamic numerical models. The difficulties of this procedure (not just numerical computations) are well known (Madariaga,1976; Madariaga et al,2006) .Therefore, to understand the $\omega^{-2}$-hypothesis for the ground displacement $u(t)$ in the far-field zone, we have focused on the physically possible local features of the source-time function $f$ that can generate the omega-square behavior of $u(t)$ at a large set of receivers. We found two appropriate fragments of $f$ : first , $f$ exhibits a local inverse-square-root behavior near the rupture front where the frontal surface is piecewise smooth ( but not smooth and not rough ). To consider such fragments as a regular and stable source of the $\omega^{-2}$-behavior , we have to admit for them the multiplicity and chaotic space orientation. The second feature is 'fractal', namely, $f$ is bounded near the rupture front where the frontal surface has a slight roughness (the Hurst parameter near 1). In this case the spectrum exhibits the $\omega^{-\theta}$ behavior with the index $\theta > 2$. It is important that $\theta \approx 2$ and this index is stable relative to receiver position.

Considering the source function fragments only, it is difficult to localize the corner frequencies without specifying the source-time function as a whole. In particular, an appropriate analysis for the fractal model by Gusev is given in (Molchan, 2015). Independently, the selected fragments can be useful for understanding the $\omega^{-2}$ behavior in the kinematic source models.

There are many other formal features of $f$ to get $\omega^{-\theta}$ - spectrum behavior with $1 < \theta < 2$ (see e.g. (29)). We can use the $\omega^{-2}$-hypothesis to reject such features as non-physical. However, to justify the $\omega^{-2}$ hypothesis we must act conversely using some (probably no local) physical restrictions. In this connection, it is unclear why the slightly rough fragments of type (26) with the parameter $\beta = 1$ (Gusev model) can generate the omega-square behavior, while this is impossible with $\beta = 1/2$ , i.e. when the time source function has a local inverse-square-root behavior near the rupture front.

.

# Appendix

## A1. Auxiliary statements (Fedoryuk, 1987)

**The Erdelyi lemma**.

If $f(t)$ is a smooth function on the semi-axis $t > o$ and $f(t) = 0$ for $t > a > 0$, then for $\beta > 0$

$$\int_0^\infty e^{i\omega t} t^{\beta-1} f(t) dt = \omega^{-\beta} f(0) \Gamma(\beta) e^{i\pi\beta/2} + \sum_1^N a_k \omega^{-(\beta+k)} + \delta_N(\omega), \omega \to \infty \qquad (A1)$$

where $\delta_N(\omega) < C\omega^{-[\beta+N]} \max\left|(d/dt)^{[\beta+N]} f(t)\right|$ and $[a]$ is the integral part of $a$.

**The method of stationary phase.**

Assume that $\Omega$ is a bounded domain in $R^n$ with a smooth boundary $\partial\Omega$, $f$ and $S$ are smooth functions in $R^n$, and $f = 0$ outside of $\Omega$. Suppose there is a unique inner point $\mathbf{x}_0$ of $\Omega$ such that $\mathbf{x}_0$ is a stationary point of $S$ and the total curvature of $S$ at $\mathbf{x}_0$, $K(\mathbf{x}_0)$, is nonzero. Then

$$\int_\Omega f(\mathbf{x}) e^{i\omega S(\mathbf{x})} d\mathbf{x} \approx a\omega^{-n/2} e^{i\omega S(\mathbf{x}_0)} (f(\mathbf{x}_0) + O(\omega^{-1})) \qquad (A2)$$

where

$$a = (2\pi)^{n/2} \left|K(\mathbf{x}_0)\right|^{-1/2} \exp(i\pi \operatorname{sgn} K(\mathbf{x}_0)/4). \qquad (A3)$$

If $\mathbf{x}_0$ does not exist, the right part of (A2) is negligibly small.

## A2. Statement from Section 2.2: the proof

Consider the model $f(\mathbf{g},t) = (t - t_r(\mathbf{g}))_+^{\beta-1} V(\mathbf{g},t)$ in a vicinity $O(\mathbf{x}_0)$ of $\mathbf{x}_0 \in \partial\Omega_f$ with smooth functions $t_r(\mathbf{g}), V(\mathbf{g},t)$. The contribution of $\mathbf{x}_0$ into the asymptotics of $\hat{u}(\omega)$ is

$$\hat{u}(\omega|\mathbf{x}_0) = \int e^{i\omega(t + <\mathbf{g}, \tilde{\mathbf{p}}>)} f(\mathbf{g},t)\varphi(\mathbf{g},t)|d\mathbf{g}|dt \qquad (A4)$$

where $\tilde{\mathbf{p}} = -\gamma_\Sigma/c$, $\varphi = \tilde{\varphi}(\mathbf{g})\hat{\varphi}(t)$ is a standard smooth function concentrated near $\mathbf{x}_0$, $|d\mathbf{g}| = dg_1 dg_2$. Using the new variable $\tau = t - t_r(\mathbf{g})$, one has

$$\hat{u}(\omega|\mathbf{x}_0) = \int e^{i\omega(t_r(\mathbf{g}) + <\mathbf{g}, \tilde{\mathbf{p}}>)} J(\omega|\mathbf{g})\tilde{\varphi}(\mathbf{g})|d\mathbf{g}|, \qquad (A5)$$

where

$$J(\omega|\mathbf{g}) = \int_0^\infty e^{i\omega\tau} \tau^{\beta-1} \tilde{V}(\mathbf{g}, \tau + t_r(\mathbf{g})) d\tau \qquad (A6)$$

and $\tilde{V}(\mathbf{g},t) = V(\mathbf{g},t)\hat{\varphi}(t)$. By the Erdelyi lemma (A1), one has

$$J(\omega|\mathbf{g}) = C_\beta \omega^{-\beta}[\tilde{V}(\mathbf{g}, t_r(\mathbf{g})) + i\beta\omega^{-1}\partial/\partial t \tilde{V}(g, t_r(g))] + \delta(\omega), \qquad (A7)$$

$$\delta(\omega) < c\,\omega^{-[\beta+2]}\max_{O(x_0)}\left|(\partial/\partial t)^2 V(\mathbf{g},t)\right|\ . \tag{A8}$$

Combining (A5) and (A7), one has

$$\hat{u}(\omega|\mathbf{x_0}) \approx \int e^{i\omega(t_r(\mathbf{g})+<\mathbf{g},\widetilde{\mathbf{p}}>)}[\widetilde{V}(\mathbf{g},t_r(\mathbf{g}))+i\beta\omega^{-1}\partial/\partial t\widetilde{V}(g,t_r(g))]\widetilde{p}(g)|d\mathbf{g}|C_\beta\omega^{-\beta}+O(\omega^{-[2+\beta]})\ . \tag{A9}$$

Suppose that the function $t_a(\mathbf{g})=t_0-<\mathbf{g},\boldsymbol{\gamma_\Sigma}/c>+t_r(\mathbf{g})$ has the unique stationary point

$\mathbf{g_\gamma}$ in $O(\mathbf{g_0})$ and the total curvature of $t_a(\mathbf{g})$ at this point is nonzero, $K_\gamma \neq 0$. Then the method of

stationary phase ( A2) gives

$$\hat{u}(\omega\mid\mathbf{x_0})\ \approx C_\beta\omega^{-(1+\beta)}2\pi\left|K_\gamma\right|^{-1/2}V(\mathbf{g_\gamma},t_r(\mathbf{g_\gamma}))\varepsilon+O(\omega^{-[2+\beta]})\ ,\ |\varepsilon|=1 \tag{A10}$$

The statement from Section 2.2 is related to the case .

If the stationary point does not exist and the $V(\mathbf{g},t)$, $t_r(\mathbf{g})$ are smooth , the method of stationary phase  implies $\hat{u}(\omega\mid\mathbf{x_0})=o(\omega^{-2})$.

## A3. Proof of (15).

Let us continue (A9), assuming that $t_r(\mathbf{g})$ is piecewise smooth. By supposition (see Section

**2.6**), $t_r(\mathbf{g})$ is represented by two smooth surfaces $t_r^\pm(\mathbf{g})$, $g \in O^\pm(\mathbf{g_0})$ in a vicinity of $\mathbf{x_0} \in \partial\Omega_f$.

These surfaces are intersected along a smooth 1-D curve $L$; the orthogonal projection of $L$ onto

the fault plane, $\hat{L}$, is given by the smooth relation $g_2=\psi(g_1)$. The integral term in (A9) is

represented as follows:

$$\hat{u}(\omega\mid\mathbf{x_0})\approx C_\beta\omega^{-\beta}e^{-i\omega t_0}(\int_{O^+(\mathbf{g_0})}+\int_{O^-(\mathbf{g_0})})e^{i\omega t_a(\mathbf{g})}\phi(\mathbf{g})|d\mathbf{g}| \tag{A11}$$

where $t_a(\mathbf{g})=t_0-<\mathbf{g},\boldsymbol{\gamma_\Sigma}/c>+t_r^\pm(\mathbf{g}|\mathbf{x_0})$, $\phi(\mathbf{g})$ are smooth and $\nabla t_a(\mathbf{g})\neq 0$ in $O^\pm(\mathbf{g_0})$ (see Section

**2.3**). Therefore the asymptotics of (A11) is determined by the common boundary of domains $O^\pm(\mathbf{g_0})$, $\hat{L}$. More precisely, it is determined by a stationary point $\widetilde{\mathbf{g}}_\gamma$ of

$t_a(\mathbf{g})$ considered as a function on $\hat{L}$. If the point $\widetilde{\mathbf{g}}_\gamma$ exists, then (see e.g., Fedoryuk, 1987)

$$\int_{O^\pm(\mathbf{g_0})}e^{i\omega t_a(\mathbf{g})}\phi(\mathbf{g})|d\mathbf{g}|\approx\pm\omega^{-3/2}i\sqrt{2\pi}e^{i\omega t_a(\widetilde{\mathbf{g}}_\gamma)+i\pi/4 s}\left|\widetilde{K}_\gamma^\pm\right|^{-1/2}(\partial/\partial g_2 t_a^\pm(\widetilde{\mathbf{g}}_\gamma))^{-1}(\phi(\widetilde{\mathbf{g}}_\gamma)+O(\omega^{-1})), \tag{A12}$$

where $\widetilde{K}_\gamma^\pm=[t_a^\pm(g_1,\psi(g_1))]''$ and $s^\pm=\mathrm{sgn}\,\widetilde{K}_\gamma^\pm$. Setting $\beta=1/2$ and combining (A11) and (A12), we get $\hat{u}(\omega\mid\mathbf{x_0})=O(\omega^{-2})$. This result holds, provided the conditional stationary point $\widetilde{\mathbf{g}}_\gamma$ of $t_a(\mathbf{g})$ exists and is regular ($\widetilde{K}_\gamma\neq 0$). The lack of $\widetilde{\mathbf{g}}_\gamma$ implies $\hat{u}(\omega\mid\mathbf{x_0})=0(\omega^{-2})$.